\documentclass[oldversion]{aa}

\usepackage{epsfig}
\usepackage{graphics}
\usepackage{float}
\usepackage{amsmath}
\usepackage{multirow}
\usepackage{longtable}
\usepackage{rotate}
\usepackage{array}
\usepackage{subfigure}
\def\kms{\ifmmode{\rm km\,s^{-1}}\else\hbox{$\rm km\,s^{-1}$}\fi}

\setlongtables

\begin{document}

\title{O-star mass-loss rates at low metallicity}

\author{L.B.Lucy}
\offprints{L.B.Lucy}
\institute{Astrophysics Group, Blackett Laboratory, Imperial College 
London, Prince Consort Road, London SW7 2AZ, UK}
\date{Received ; Accepted }

\abstract{Mass fluxes $J$ are computed for the extragalactic O stars 
investigated by
Tramper et al. (2011; TSKK). For one early-type O star, computed and
observed rates agree within errors. However, for 
two late-type O stars, theoretical mass-loss rates underpredict observed 
rates by
$\sim 1.6$ dex, far exceeding observational errors. 
A likely cause
of the discrepancy is overestimated observed rates due to the neglect of
wind-clumping. A less likely but   
intriguing possibility is that, in observing O stars with
$Z/Z_{\sun} \sim 1/7$, TSKK have serendipitously discovered 
an additional mass-loss mechanism not evident in the spectra of
Galactic O stars with powerful radiation-driven winds.
Constraints on this unknown mechanism are discussed.\\
  In establishing that the discrepancies, if real, are inescapable
for purely radiation-driven winds, failed searches for high-$J$ 
solutions are reported and the importance of a numerical technique
that cannot spuriously create or destroy momentum stressed.\\
  The $Z$-dependences of the computed rates for $Z/Z_{\sun} \in (1/30, \: 2)$ 
show significant departures from
a single power law, and these are attributed to curve-of-growth effects in the
differentially expanding reversing layers. The best-fitting power-law exponents 
range from 0.68-0.97.
}

\authorrunning{Lucy}
\titlerunning{O-star mass-loss rates}
\maketitle

\section{Introduction}

By observing O stars in nearby dwarf galaxies with metallicities ($Z$) below
that of the
Small Magellanic Cloud, Tramper et al. (2011; TSKK) have extended the range
of $Z$
over which O-star winds have been detected and analysed. The resulting
mass-loss rates ($\Phi$) for $Z/Z_{\sun} \sim 1/7$ are surprisingly high,
leading the authors to question the Vink et al. (2001) scaling law
$\Phi \propto Z^{\alpha}$ with $\alpha = 0.69 \pm 0.10$.

Because of this result's importance for the later evolutionary phases of 
massive stars and perhaps also for the theory of stellar winds,
models computed with the TSKK stars'
effective temperatures ($T_{\rm eff}$) and surface gravities ($g$) are of 
interest in order
to quantify discrepancies without relying on scaling laws. In addition,
the validity of power-law scaling merits investigation. Accordingly,
in this paper, mass fluxes $J = \Phi/ 4 \pi R^{2}$ are computed  
for a wide range of $Z$
at several points in $(T_{\rm eff},g)$-space.

Throughout this paper, the units of $J$ and $\Phi$ are gm s$^{-1}$ cm$^{-2}$ 
and $\sun \: yr^{-1}$, respectively. In these units,
\begin{equation}
  log \Phi = log J + 2 \: log \: R/R_{\sun} - 3.015 
\end{equation}

\section{Mass Fluxes}

In this section, the assumptions and procedures adopted when computing $J$
from the theory of moving reversing layers (RL's) are briefly reviewed, as are
the curve-of-growth (CoG) effects that control $J$'s dependence
on $Z$.
 
\subsection{Assumptions} 

The fundamental assumption is that the $\Phi$ of a stationary, 
radiatively-driven wind
is determined by the constraint of regularity at the sonic point - i.e.,
that the transonic flow makes a smooth (analytic) transition from sub-
to supersonic velocities. This assumption dates back to Lucy \& Solomon's
(1970; LS70) primitive theory of dynamic RL's and retains its
role in updated versions (Lucy 2007, 2010a,b; L07, L10a,b).

Additional assumptions are: 1) plane-parallel, isothermal flow;
2) continuum emission occurs only at the lower boundary of the 
Schuster-Schwarzschild RL and has the frequency distribution of the
TLUSTY model atmosphere with the prescribed 
$T_{\rm eff}, g$ and
$Z$ (Lanz \& Hubeny 2003) ; 3) line formation occurs by non-coherent 
scattering with
complete redistribution in the fluid frame; 4) line broadening is due 
radiation damping and Doppler broadening with 
microturbulent velocity $v_{t} = 10$km s$^{-1}$; 
5) the RL's upper boundary condition of no incoming radiation 
is imposed at velocity $v = 5a$, where $a$ is the isothermal sound speed.
   
\subsection{Procedures} 

This multi-line transfer problem in a differentially expanding RL
is treated with a Monte Carlo (MC) technique. This has the considerable
merit of coding
simplicity but has the drawback that $\tilde{g}_{\ell}$, the MC estimate
for the contribution of lines to
the radiative acceleration, is not analytic. This poses the problem of
how to determine the eigenvalue $J$ from the regularity condition
\begin{equation}
  g_{\rm eff} = g_{*} - g_{\ell} = 0  \;\;\: at \:\;\; v = a
\end{equation}
where $g_{*} = g - g_{e}$ and $g_{e}$ is the radiative acceleration due to
electron scattering.

The procedure developed in L07 and L10b replaces this {\em point}
constraint by an {\em integral} constraint $Q_{1,2} = 0$, which is 
here conveniently rewritten as
\begin{equation}
  \Delta_{1,2} =
  \int_{x_{1}}^{x_{2}} g_{\ell} \: dx = \left|\: \frac{1}{2} v^{2} 
             - a^{2} \ln v + g_{*} x \: \right|_{x_{1}}^{x_{2}}
\end{equation}
where the heights $x_{1}$ and $x_{2}$ correspond to  
$v/a \approx 1/2$ and $\approx 2$, respectively. The interval ($x_{1},x_{2}$)
includes the sonic point but excludes zones 
where the solution is affected by inexact boundary conditions. 

A MC simulation with ${\cal N}$ indivisible photon packets is carried 
out in a model RL derived from the trial
solution: $g_{\ell} = g^{\ell}_{i}$ at $v_{i}$ for 
$i=1,2, \ldots ,k, \ldots ,I$ 
with $g^{\ell}_{k} = g_{*}$ at $v_{k} = a$. The resulting estimates 
$\tilde{g}_{\ell}$ then allow the
left-hand side of Eq.(2) to be evaluated numerically, thereby testing if
the equation is satisfied. With trial-and-error adjustments of
$g^{\ell}_{i}$ for $i \ne k$, sequences of such simulations
allow the eigenvalue $J$ to be determined (L10a,b).

This numerical procedure achieves {\em consistency} with the posed
integro-differential problem in the limits $I \rightarrow \infty$  
and ${\cal N} \rightarrow \infty$, provided that the iterative adjustments of
$J$ and the $g^{\ell}_{i}$ have resulted in exact agreement between 
$\tilde{g}_{\ell}$ and $g^{\ell}_{i}$ throughout the RL. 

The main weakness at present of this technique is its reliance on 
trial-and-error adjustment.
In consequence, each solution demands a large commitment of the 
investigator's time. 
A reliable iterative algorithm is needed if large numbers of
solutions are required.

Because the MC quanta are {\em indivisible}, none are created or 
destroyed {\em within} the RL. Accordingly, this MC technique does not 
spuriously create or destroy momentum. The sonic-point momentum
flux $Ja$ therefore derives from identifiable physical processes.

\subsection{Curve-of-growth effects} 

Classically, the CoG  refers to {\em static} RL's and
relates the emergent line profile
and its equivalent width to $N$, the absorbing ion's column 
density above the photosphere. However,
the physical effects underlying the CoG 
are also relevant (LS70) to the $J$ that can be radiatively-driven through the
sonic point since 
$g_{\ell}$ at $v = a$ depends on the co-moving radiation field,
and this reflects line formation throughout the RL, including the quasi-static
layers close to the photosphere.

For simplicity, consider an outflow driven by {\em one} line from an 
atmosphere with continuum absorption due only to H
and imagine varying $Z$ with fixed $T_{\rm eff}, g$. 
We follow the discussion in LS70
for the $\lambda1548\AA$ component of the C $\:${\sc iv} doublet
when the dominant ion is C $\:${\sc iii}. For vanishingly small $Z$ - i.e.,
negligible line formation,
$g_{\ell}^{\dag} \propto n_{1}^{\dag}$, the ground state number density at 
$v^{\dag} = a$. On the assumption of an
ionization balance between radiative recombinations and photoionizations,
$n_{1}^{\dag} \propto n_{C}^{\dag}/n_{e}^{\dag} \propto Z/\rho^{\dag}$. 
Accordingly, the sonic point constraint
$g_{\ell}^{\dag} = g_{*}$ is obeyed at density $\rho^{\dag} \propto Z$, whence
$J =  \rho^{\dag} v^{\dag} \propto Z$. 

In this weak-line limit, $J \propto Z$ because of increasing
$N$. This effect is seen in Fig.2 of LS70 at 
$log T_{\rm eff} \approx 4.55$ and at $\approx 4.25$, where increasing     
$N$ is due to shifting ionization balance as $T_{\rm eff}$ increases.

When increasing $Z$ invalidates the weak-line limit, line formation reduces the
co-moving flux seen by ions at $v \sim a$ and so $J(Z)$ flattens,
reaching a maximum at $J \sim J_{*} = \sigma T_{\rm eff}^{4}/c^{2}$, when all
the photon momentum accessible with {\em one} line is
converted into momentum of the transonic flow (LS70). In terms of the 
dimensionless rate
$\phi = \Phi c^{2}/ L$, this maximum corresponds to $\phi \sim 1$.
  
When increasing $Z$ brings about the strong-line limit, corresponding to a
line with a saturated Doppler core and well-developed damping wings,
the photon momentum available at $v \sim a$ is drastcally reduced, as therefore
is $J$. This effect is seen in Fig.2 of LS70 at $log T_{\rm eff} \approx 4.5$ 
In this strong-line limit, the accessible photon momentum in the continuum
is transferred to quasi-stationary gas ($v^{2} \ll a^{2}$) deep in the RL
where it merely increases $H_{\rho}$, the local scale height.

With more than one line, the above CoG effects are superposed. However,
in regions of high line-density, effects of overlapping line formation come
into play. These defy simple interpretation and are the main reason
for adopting MC methods.

\section{Models with varying Z} 

In this section, models are computed at 
selected points in $(T_{\rm eff},g,Z)$-space.

\subsection{Parameters}

Because the RL's require input from the OSTAR2002 models of
Lanz \& Hubeny (2003), their grid somewhat restricts the parameter choices
made here. We adopt the TLUSTY labelling scheme, so that, for example,
model Gt400g300 has $T_{\rm eff} = 40.0$kK, $log g = 3$ and Galactic
metallicity. In order to test scaling laws over the full range
of metallicites found
in the local Universe, a sequence of models 
with letter codes C,G,L,S,T and V are computed at each selected
point ($T_{\rm eff},g$).
These correspond, respectively, to $Z/Z_{\sun} = 2, \:1,\: 1/2,\: 1/5,\: 1/10$
and $1/30$, a range of 60 in metallicity.

With regard to $T_{\rm eff}$ and $g$, points in the OSTAR2002 grid
are chosen to match the best-fit parameters in Table 2 of TSKK.
The chosen points and the stars matched within errors are:           

   t475g375 (IC 1613-A13)

   t425g375

   t350g350 (IC 1613-A15; IC 1613-C9; NGC 3109-20)  

   t300g325 (WLM-A11; IC 1613-B11)\\
The t425g375 sequence is included to avoid a large gap in $T_{\rm eff}$.

All models have $v_{t} = 10$km s$^{-1}$. This canonical value is consistent 
with the highly uncertain TSKK estimates. Note that well-determined
values of $v_{t}$ for Galactic O stars
(e.g., Bouret et al. 2005) derive from 
UV spectra, which are not available for the TSKK stars.

\subsection{Model sequencies}

Values of $J$ for the four sequencies are given in Table 1.
The additional tabulated quantities are:\\ 

{\em Col 4}: $\: \phi$, the mass-loss rate in units of $L/c^{2}$.\\

Since $\phi$ has a maximum value $\sim 1$ with one driving line (LS70), 
$\phi$ is the equivalent number of effective non-overlapping lines.\\   

{\em Col. 5}: $\: \eta$, the percentage of the photon packets escaping
across the RL's upper boundary that have {\em not} undergone a
line scattering event within the RL.\\

Since the MC quanta emitted at the lower boundary sample the 
continuum flux distribution of the TLUSTY model, a high value of $\eta$
indicates that much of this emission occurs in relatively line-free frequency
intervals. Conversely, a low value indicates effective line-blocking of
the star's continuum.\\ 

{\em Col. 6}: $\: \zeta$, the percentage of $\Delta_{1,2}$ - see Eq.(3) -
contributed by lines of Fe and Ni.\\

This quantity measures the importance of Fe and Ni relative to
C,N,O,Ne,Si and S in accelerating matter through the sonic point. Note that
the MC estimator for $g_{\ell}$ - Eq.(10) in L07 - readily allows the 
contributions of individual lines to be extracted.

\begin{table}

\caption{Computed mass fluxes $J$(gm s$^{-1}$ cm$^{-2}$).}

\label{table:1}

\centering

\begin{tabular}{c c c c c c}

\hline\hline

 Seq.   & $Z/Z_{\sun}$ & $\log J$ & $\phi$ & $\eta (\%)$ & $\zeta(\%)$  \\

\hline
\hline

              & 2    &   -3.86  &  429.8  & 37.8 & 80.0 \\

              & 1    &   -4.38  &  129.8  & 47.2 & 81.5 \\

 $t475g375$   & 1/2  &   -4.66  &   68.1  & 52.4 & 80.3 \\

              & 1/5  &   -4.98  &   32.6  & 58.6 & 69.9 \\

              & 1/10 &   -5.17  &   21.1  & 58.0 & 58.7 \\

              & 1/30 &   -5.72  &    5.9  & 73.9 & 23.6 \\

\cline{1-6}

              & 2    &   -5.12  &   36.9  & 62.8 & 89.4 \\

              & 1    &   -5.13  &   36.0  & 65.4 & 86.4 \\

 $t425g375$   & 1/2  &   -5.58  &   12.8  & 69.8 & 64.6 \\

              & 1/5  &   -5.93  &    5.7  & 74.5 & 38.7 \\

              & 1/10 &   -6.21  &    3.0  & 78.5 & 21.4 \\

              & 1/30 &   -6.60  &    1.2  & 84.6 &  9.9 \\

\cline{1-6}

              & 2    &   -6.06  &    9.2  & 61.0 & 64.6 \\

              & 1    &   -6.08  &    8.8  & 66.1 & 53.3 \\

 $t350g350$   & 1/2  &   -6.10  &    8.4  & 70.0 & 50.6 \\

              & 1/5  &   -6.78  &    1.8  & 76.4 & 23.2 \\

              & 1/10 &   -6.76  &    1.8  & 79.6 & 14.6 \\

              & 1/30 &   -7.17  &    0.7  & 81.8 &  5.7 \\

\cline{1-6}

              & 2    &   -6.13  &   14.5  & 59.3 & 50.4 \\

              & 1    &   -6.50  &    6.2  & 64.6 & 33.2 \\

 $t300g325$   & 1/2  &   -6.62  &    4.7  & 69.7 & 21.8 \\

              & 1/5  &   -7.22  &    1.2  & 76.2 &  5.2 \\

              & 1/10 &   -7.18  &    1.3  & 79.9 &  4.1 \\

              & 1/30 &   -7.46  &    0.7  & 85.9 &  2.1 \\

\hline
\hline

\end{tabular}

\end{table}

\subsection{Dependence of $J$ on metallicity}

In Fig.1, the $J$'s given in Table 1 are plotted against $Z/Z_{\sun}$. 
These $log-log$ plots show significant departures from linearity and even
include departures from the expected monotonic decline with decreasing Z.

The quantities $\eta$ and $\zeta$ provide a qualitative understanding of the
$J$'s dependence on $Z$. In the main, we see that $J$ increases with
$Z$ but that $\eta$ decreases. Thus, the enhanced $J$ and $\phi$ can be 
attributed to the shrinking of line-free frequency intervals 
as weak lines become significant absorbers.  

The importance of CoG effects is illustrated by $\zeta$. The spectra of
Fe and Ni ions have numerous weak permitted lines arising from metastable
levels. In contrast, the line-driving contributed by light and intermediate
mass elements is predominantly from strong ground-state resonance transitions.
As $Z$ increases, these ground-state transitions saturate, and their 
momentum tranfers then
occur deep in the RL, increasing $H_{\rho}$ but contributing little
to $J$. Because of this effect, the relative contribution of Fe and Ni
to $\Delta_{1,2}$ increases dramatically with $Z$.

\begin{figure}
\vspace{8.2cm}
\includegraphics{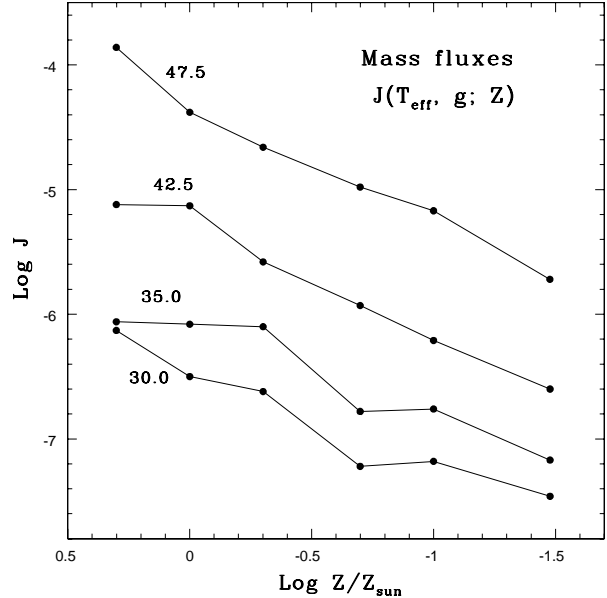}
\caption{Mass fluxes $J$ from Table 1 plotted against metallicity
$Z/Z_{\sun}$. The curves are labelled with the sequencies' $T_{\rm eff}(kK)$.}  
\end{figure}

\subsection{Power-law exponents $\alpha$}

Conventionally, the Z-dependence of $\Phi$ is approximated as a power
law, so that
\begin{equation}
log J = log J_{G} + \alpha \: log Z/Z_{\sun} 
\end{equation}
Although Fig.1 shows significant departures from
this behaviour, least squares fits of Eq.(4) to the data have been carried
out and are given in Table 2. The intercepts $log J_{G}$ are not reported since
they do not supersede the individual $J$-values for the 
$G$-models.

In view of the significant departures from power-law behaviour, the 3-D
function $J(T_{\rm eff}, g ; Z)$ will eventually have to be tabulated 
so that the evolution of massive stars can be followed as a
function of initial metallicity.

\begin{table}

\caption{Power law exponents $\alpha$.}

\label{table:1}

\centering

\begin{tabular}{c c c}

\hline\hline

 Seq.   & $\alpha \pm \sigma_{\alpha}$ & $\sigma_{log J}$ \\

\hline
\hline

 $t475g375$ &  0.97 $\pm$ 0.07  &   0.10  \\

 $t425g375$  &  0.89 $\pm$ 0.07  &   0.10  \\

 $t350g350$ &  0.68 $\pm$ 0.11  &   0.17  \\

 $t300g325$ &  0.75 $\pm$ 0.10  &   0.15  \\

\hline
\hline

\end{tabular}

\end{table}

\section{Comparisons with observed $\Phi$'s}

In this section, the $\Phi$'s derived by TSKK for three of their
stars are compared to the predictions of Sect.3. 
The stars chosen are the one with high $T_{\rm eff}$ and the two late-type
O stars with the most accurate $\Phi$'s. 

\subsection{IC1613-A13}

For this star, TSKK estimate $T_{\rm eff}(kK) = 47.6^{+4.73}_{-4.95}$ and
$log g = 3.73^{+0.13}_{-0.22}$ and so the appropriate sequence from Table 1 is
$t475g375$. The metallicity of IC1613 is $Z/Z_{\sun} \approx 1/7$,  
which is bracketed by the models with $Z/Z_{\sun} = 1/5$ and $1/10$.
Logarithmic interpolation gives $J = -5.07$ dex. Accordingly,
with $R/R_{\sun} = 11.4$ from TSKK, 
$\Phi = 4 \pi R^{2} J = -5.97$ dex. This is to be
compared with the TSKK estimate $\Phi = -6.26^{+0.45}_{-0.50}$ dex. 
Thus theory
overpredicts observation by 0.29 dex, but this is well within observational
uncertainty and so there is no conflict.

\subsection{NGC 3109-20}

For this star, TSKK derive $T_{\rm eff}(kK) = 34.2^{+4.70}_{-3.05}$ and
$log g = 3.48^{+0.37}_{-0.40}$ and so the appropriate sequence is
$t350g350$. The metallicity of NGC 3109 is $Z/Z_{\sun} \approx 1/7$.  
Logarithmic interpolation gives $J = -6.77$ dex. Accordingly,
with $R/R_{\sun} = 24.7$ from TSKK,  
$\Phi = -7.00$ dex. This is to be
compared to the TSKK estimate that $\Phi = -5.41^{+0.25}_{-0.35}$ dex. 
Thus theory
underpredicts observation by 1.59 dex, far beyond the
observational uncertainty.  

In dimensionless terms, the TSKK rate is $\phi = 76$, indicating
the number of effective driving lines required to accelerates the corresponding
$J$ through the sonic point. Inspection of Table 1 shows that $\phi = 76$
is only achieved for $T_{\rm eff} \ga 45.0$ and $Z/Z_{\sun} \ga 1/2$,
a domain excluded by TSKK's spectroscopic analysis. Moreover, the interpolated
$\eta$ for this model is $\approx 78$ per cent, showing that there are 
extensive
frequency intervals not contributing to line driving of the transonic flow.

\subsection{WLM-A11}

For this star, TSKK derive $T_{\rm eff}(kK) = 29.7^{+2.45}_{-2.75}$ and
$log g = 3.25^{+0.29}_{-0.19}$ and so the appropriate sequence is
$t300g325$. The metallicity of WLM is again $Z/Z_{\sun} \approx 1/7$.  
In this case, interpolation gives $J = -7.20$ dex. Accordingly,
with $R/R_{\sun} = 29.8$ from TSKK, 
$\Phi = -7.27$ dex. This is to be
compared with the TSKK estimate that $\Phi = -5.56^{+0.20}_{-0.30}$ 
dex. Thus theory
underpredicts observation by 1.71 dex, far outside the
observational uncertainty.

In dimensionless terms, the TSKK rate is $\phi = 66$
Thus, as for NGC 3109-20, this  
is only achieved for $T_{\rm eff} \ga 45.0$ and 
$Z/Z_{\sun} \ga 1/2$,
a domain excluded by TSKK's spectroscopic analysis. 
The interpolated $\eta$ for this model is also $\approx 78$ per cent, thus 
showing again that there are extensive
frequency intervals that cannot be tapped to sustain a high $J$.
  
\section{Discussion}

In this section, the severe conflict between theory and observation
for the two late-type O stars is discussed. For clarity, the two extremes
are considered.   

\subsection{Observed rates in error}

The possible causes of overestimation of observed rates are discussed
in detail by TSKK. As they fully recognize, a likely cause is wind clumping. 
Their estimates of
$\Phi$ derive from the partial filling-in of the 
He$\:${\sc ii} $\lambda4686\stackrel{o}{A}$ and H$\alpha$ absorption
lines by wind emission - see their Fig.2. Now, it has long been suspected that
the instability of line-driven winds will lead to a clumpy outflow and hence
enhanced emission in exactly these lines (Lucy 1975). Moreover, the standard
diagnostic code CMFGEN (Hillier \& Miller 1999) includes an adjustable 
clumping factor that demonstrably improves fits to observed spectra 
(e.g., Bouret et al. 2008).    
Nevertheless, the model wind (Puls et al. 2005)
that SDKK use to fit the data does not allow for clumpiness.
Accordingly, one suspects that, if clumpiness were included, their genetic 
algorithm (Mokiem et al. 2005) would find a wide range of acceptable 
solutions: from high $\Phi$'s with weak clumping to low $\Phi$'s with strong 
clumping - i.e., degeneracy with respect to these two parameters.

In view of the limited information content concerning winds in the
optical spectra of O stars, the $\Phi$'s of the TSKK stars are likely to 
remain highly
uncertain until UV spectra can be obtained. However, in the meantime,
as a control experiment, the TSKK methodology should be applied to optical
spectra of Galactic O stars that already have analysed UV spectra.
If the $\Phi$'s of Galactic O stars with similar spectral types to NGC 3109-20
and WLM-A11 can thus be derived with reasonable accuracy purely
from their optical spectra, then the high TSKK $\Phi$'s 
for these low-Z extragalactic O stars would be more convincing.

\subsection{Theoretical rates in error}

The accuracy of the eigenvalues $J$ is discussed
in Sect. 4.2 of L10b, from which it follows that errors $\sim 1.6$ dex would
appear to be excluded. However,
there is no proof that the eigenvalue $J$
is unique. Accordingly, a search for a second, high-$J$ solution
has been carried out for St350g350 ($Z = Z_{\sun}/5$). Thus,
as described in Sect. 2.2,
the vector $g^{\ell}_{i}$ for $i \ne k$ is iteratively corrected with the
aim of achieving dynamical consistency by matching the MC estimates  
$\tilde{g}_{\ell}$. But now $J$ is {\em fixed} at -5.19 dex, a value obtained 
by adding the discrepancy of 1.59 dex found in Sect. 4.2 for NGC 3109-20 to 
the eigenvalue $J = -6.78$ dex from Table 1 (cf. Sect. 3.3 in L10a).

Fig. 2 illustrates the limited degree of consistency achieved when the
search was abandoned. This may be compared with Fig.2 in L10b,
where the successful solution for $Gt400g375$ is plotted. In that case,
$J$ was {\em not} fixed but instead was iteratively corrected to obtain 
$\Delta_{1,2} = 0$.    

Fig. 2 shows that dynamical consistency is achievable for $v/a \la 0.6$
but fails at higher velocities. Evidently, the constraint that 
the dynamical RL satisfies the regularity constraint (Eq. [2]) precludes  
adjustment into agreement with $\tilde{g}_{\ell}$ as the
sonic point is approached.
Accordingly, the inescapable conclusion is that a second solution with
$\Delta log J \sim 1.6$ does not exist.

An identical experiment for $St300g325$ with $J$ fixed at 
$-7.22 + 1.71 = -5.11$ dex also ended in a failure similar to Fig.2.

The MC code cannot match these {\em imposed} sonic point
momentum fluxes $Ja$ because there are no physical processes available
to provide these fluxes {\em and} the use of indivisible MC quanta precludes
the spurious creation of momentum (Sect. 2.2).

\begin{figure}
\vspace{8.2cm}
\includegraphics{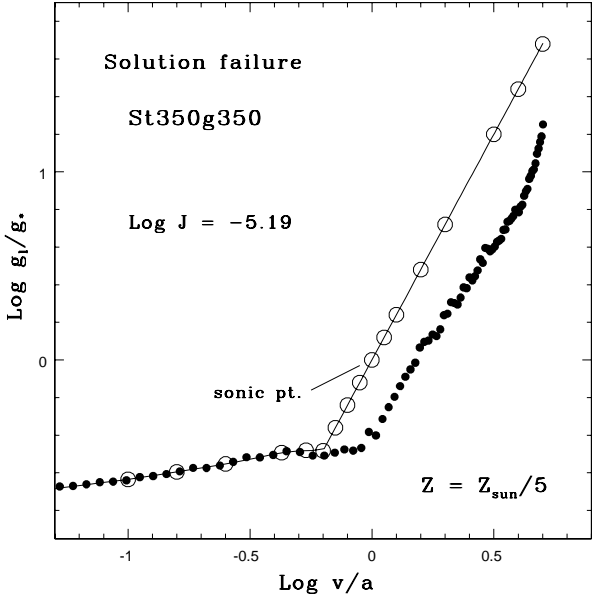}
\caption{Failed solution for $St350g350$ when $J = -5.19$ dex.
The open circles are the $g_{i}^{\ell}$ for $i=1,2, \ldots ,k, \ldots ,I$ 
from which which the model RL is derived. The filled 
circles are the MC estimates $\tilde{g}_{\ell}$. The sonic point ($i=k$) is
indicated.}
\end{figure}

\subsection{An additional mass-loss mechanism?} 

If the $\Phi$ estimates of TSKK are accurate, then the above 
elimination of high-$J$ solutions implies 
that an additional acceleration mechanism is operating besides radiative 
driving. This is perhaps an unwelcome possibility,
but where better to discover an additional mechanism than
in the spectra O stars with low-$Z$'s?

For steady flow, Fig.2 shows that this unknown mechanism must provide
an outward force per gm of $\sim 0.4g$ at $v = a$ in order to make up for
$\tilde{g}_{\ell}$'s shortfall in satsfying Eq.(2).
But extra driving is not 
needed at higher velocities. For $v/a \ga 1.5$, 
Fig.2 shows that  
a net outward acceleration is provided by $\tilde{g}_{\ell}$.   
This is a consequence of matter being Doppler-shifted out of the absorption
lines formed in the RL's quasi-static layers.

The possibility that an extra mechanism may be required only in the
neighbourhood of the sonic point prompts the thought that
small-scale, stochastic expulsions may be accelerating blobs of photospheric
matter to $v/a \ga 1.5$, whereupon line-driving takes over and a high-$\Phi$
supersonic wind ensues.   

A strong constraint on this unkown mechanism is provided by the
existence of a {\em weak-wind domain} 
for Galactic O stars (Marcolino et al. 2009; M09).
From Table 3 in M09, we see that, observationally, 
this domain's extent is $T_{\rm eff}(kK)  \in (31,34)$ and 
$log g \in (3.6,4.0)$. Thus, in $(T_{\rm eff}, g)$-space, 
the two late-type O stars of TSKK are contiguous to the weak-wind domain,
whose existence is solidly-based on UV spectra and is partially understood
theoretically (L10a,b).
Evidently, the unknown mechanism does not operate in the weak-wind domain.     
Accordingly, it must be narrowly focused in parameter space, perhaps also
excluding $Z/Z_{\sun} \sim 1$.

\section{Conclusion}

The aim of this paper has been to test 
the theory of moving RL's at metallicities other than Galactic.
The stimulus was the high empirical $\Phi$'s reported recently by TSKK for
extragalactic O stars with $Z/Z_{\sun} \approx 1/7$ and the doubts they
raised about the Vink et al. (2001) scaling law.

Although agreement within errors is found at $T_{\rm eff}(kK) \sim 47.5$,
discrepacies $\sim 1.6-1.7$ dex are found for two late type O stars
with $T_{\rm eff}(kK) \sim 30 - 35$. If these discrepancies are not
due to wind clumping, an effect not included by TSKK, then an additional
mass-loss mechanism is necessary, perhaps
operating only at low velocities and in a restricted part of
$(T_{\rm eff}, g, Z)$-space (Sect. 5.3). What this mechanism might be is
an open question. 

Most probably, wind-clumping is the solution to these
discrepancies, but the proof may not come until we can obtain spectra
in the UV.

\end{document}